# Development of an electron impact ion source with high ionization efficiency for future planetary missions

Oya Kawashima[a*], Naho Yanase[a], Yoshihisa Okitsu[a], Masafumi Hirahara[b], Yoshifumi Saito[a,c], Yuzuru Karouji[c], Naoki Yamamoto[a,c], Shoichiro Yokota[d], and Satoshi Kasahara[a]

[a] *The University of Tokyo, 7-3-1 Hongo, Bunkyo, Tokyo, 113-0033, Japan*

[b] *Nagoya University, Nagoya, 464-8601, Japan*

[c] *Institute of Space and Astronautical Science (ISAS), Japan Aerospace Exploration Agency (JAXA), Sagamihara, 252-5210, Japan*

[d] *Osaka University, Toyonaka, Osaka, 560-0043, Japan*

\* Corresponding author

*E-mail address:* o.kawashima@eps.s.u-tokyo.ac.jp (Oya Kawashima)

**Abstract**

Ion sources using electron impact ionization (EI) methods have been widely accepted in mass spectrometry for planetary exploration missions because of their simplicity. Previous space-borne mass spectrometers were primarily designed with the EI method using rhenium tungsten alloy filaments, enabling up to 100–200 μA emission in typical cases. The emission level is desired to be enhanced because the sensitivity of mass spectrometers is a critical requirement for the future in situ mass spectrometry related to the measurement of trace components in planetary samples. In this study, we developed a new high-emission EI ion source using a $Y_2O_3$-coated iridium filament, which has a lower work function than rhenium tungsten alloy. The size of the ion source was 30 mm × 26 mm × 70 mm, and its weight was ~70 g. We confirmed that when consuming ~3.0 W power, the ion source records 1–2 mA electrons, which is 10 times greater than the conventional models' electron emission level. We verified the linearity of ionization efficiency and the electron current in the range of 0.1 to 1 mA, which indicates our new model increased the ionization efficiency. We conducted performance tests on the prototype with the 3.0 W heating condition, confirming a high ionization efficiency (~$10^4$ nA/Pa). In addition, we conducted endurance tests of the ion source and demonstrated the persistence of the ionization efficiency for 30 min × 100 cycles.

**Keywords**: mass spectrometry, space exploration, electron impact ionization, $Y_2O_3$ coated Iridium cathode, high electron emission

Oya Kawashima, Masafumi Hirahara, Yoshifumi Saito, Naoki Yamoto, Yuzuru Karouji, Shoichiro Yokota and Satoshi Kasahara substantially contributed to the study conceptualization. Oya Kawashima, Naho Yanase, Yoshihisa Okitsu, Yoshifumi Saito, Naoki Yamoto and Satoshi Kasahara significantly contributed to experimental demonstration, data analysis and interpretation. All authors critically reviewed and revised the manuscript draft and approved the version for submission.

1.  **Introduction**

Mass spectrometry is a powerful tool for the in-situ analysis of planetary samples in space exploration. This method has been critical in understanding the history of the planets. Because the elemental and isotopic compositions of planetary samples indicate the building blocks of the parent bodies and their later evolutions, space-borne mass spectrometers have been intensively developed and applied for solar system exploration. For example, recent achievements include the analysis of Enceladus plumes (Waite Jr et al., 2006), Martian gases (Mahaffy et al., 2013), and the coma of the 67P-CG comet (Altwegg et al., 2014).

Mass spectrometers for future missions require a higher sensitivity for the measurement of trace components in planetary samples that previous instruments could not achieve. From a technological viewpoint, ionization efficiency is one of the key parameters for obtaining high sensitivity. In most previous planetary missions, ionization sources were designed using the electron impact ionization (EI) method, in which neutral gas analytes are bombarded with a beam of energetic electrons. One main reason for its popularity is the simplicity of this technique. Electrons can be produced through thermionic emission from metal filaments using Joule heating alone. The original design of a space-borne EI ion source was described by Nier (1940), in which a tungsten filament was combined with simple electrodes. This type of filament has been widely used for several decades as cathode; in recent missions, the material has been altered to a rhenium tungsten alloy with a diameter of approximately 0.075–0.1 mm, which enables 100–200 µA emission with ~1 W of power (Niemann et al., 2002; Balsiger et al., 2007; Mahaffy et al., 2012; Arevalo Jr et al., 2015). Additionally, downsizing efforts are applied to the conventional models using the rhenium tungsten alloy (Ren et al., 2020).

The ionization efficiency of the EI ion sources is controlled by two factors: the electron emission current (i.e., the amount of bombarding electrons) and collisional electron energy. Generally, cross-sections of EI ionization for many gas species are maximal when ~70 eV electrons are bombarded (Engel, 1965). Accordingly, the electron emission current should be enhanced to measure thinner planetary components. In this study, we chose thermionic sources rather than field emission sources, considering that the electron emission current is generally lower than 100 µA for field emission sources (e.g., Krysztof 2021). We applied another cathode type with a lower work function than that of the rhenium tungsten alloy filaments to emit larger amounts of electrons at the same temperature.

In Section 2, we presented an overview of the theoretical design of the ion source and our results from the beam trajectory calculation. In Section 3, we summarized the results of the instrument's experimental and endurance tests. In Section 4, we concluded the paper with various features of the new ion source.

## 2. Theory and methods

### 2.1. Cathode selection

The material choice for the cathode is an important parameter in the design of EI ion sources. There are a variety of different types of cathodes to consider. Table 1 lists the physical properties of commercial cathode materials (Zaima et al., 1981). To obtain a high-density electron beam, materials with a low work function and high melting point are preferable, considering the Richardson equation (1):

$$J_R = AT^2 \exp\left(-\frac{\Phi}{k_B T}\right) \tag{1}$$

where $J_R$, $A$, $T$, $\Phi$, and $k_B$ denote the maximum achievable thermionic electron current density, Richardson's constant, cathode temperature, work function, and Boltzmann constant, respectively. To be emitted from a cathode into a vacuum, electrons need to overcome the energy threshold of the work function; thus, lower work-function materials enable higher electron emission at the same temperature. From another perspective, it is preferable to select lower-work-function materials because the higher temperature of the cathode heats the surrounding structure and causes a larger amount of degassing, which results in a shorter cathode lifetime. The melting point of the cathode material is also important, as its functional lifetime is shortened if the operating temperature of the cathode approaches this value. Thus, there is an upper limit to the operational temperature for each material, and therefore a high melting point is preferable for the cathode.

**Table 1.** Physical properties of various cathode materials (Data from Zaima et al., 1981).

| Material | Work function [eV] | Melting point [$10^3$ K] | Resistivity [$10^{-7}$ Ω·m] |
|---|---|---|---|
| $LaB_6$ | 2.69 | 2.80 | 1.5 (300 K) |
| $CeB_6$ | 2.73 | 2.56 | 2.9 (300 K) |
| CaO | 1.78 | 2.84 | $2.8 \times 10^7$ (1100 K) |
| SrO | 1.43 | 2.77 | $18 \times 10^7$ (1100 K) |
| BaO | 1.25 | 2.19 | $4.6 \times 10^7$ (1100 K) |
| $ThO_2$ | 2.78 | 3.46 | $0.65 \times 10^7$ (1100 K) |
| $Y_2O_3$ | 2.00 | 2.81 | $0.10 \times 10^7$ (1100 K) |
| TiC | 3.32 | 3.43 | 5.3 (300 K) |
| ZrC | 3.38 | 3.80 | 6.2 (300 K) |
| TaC | 3.61 | 4.15 | 4.1 (300 K) |
| $CeC_2$ | 2.49 | 2.81 | 6.0 (300 K) |
| W | 4.54 | 3.68 | 0.55 (300 K) |
| Ir | 5.6 | 2.72 | 0.52 (300 K) |
| Ta | 4.1 | 3.27 | 1.3 (300 K) |
| Th | 3.71 | 2.02 | 1.5 (300 K) |
| Ba | 2.11 | 1.12 | 6.0 (300 K) |
| Re | 4.55–4.59 | 3.05 | 2.9 (300K) |

Considering the work function and the melting point of the rhenium tungsten alloys, these are not the first candidates to be selected. Meanwhile, oxide materials appear to be more suitable for this purpose. However, the electrical conductivities of oxide materials are extremely low and therefore cannot be used as thermionic cathodes. In this study, we selected an oxide material as an additive coating on a conductive base-metal filament. The oxide material was heated via thermal conduction from the central metal. We chose $Y_2O_3$ for coating material because of its proven performance as a commercial ion gauge (e.g., Nakajima 2016; Rutkov and Gall, 2019). Considering the tolerance to chemically active gases, an iridium filament was selected as the base metal (Melton, 1958). Note that other set of materials with similar physical properties can also be reasonably chosen for the cathode. We manufactured a cathode with a diameter of 0.125 mm iridium filament and $Y_2O_3$ with a coverage thickness of 0.05 mm in cooperation with HORIZON Corporation, Tokyo, Japan. The $Y_2O_3$ coating over iridium filament is manufactured with electrodeposition method. We selected straightened filaments rather than conventional hairpin-shaped filaments (see Fig. 1).

In conclusion of this section, we present advantages of our cathode in comparison with a few previous studies. A BaO-based cathode was applied to the Strofio instrument in the SERENA package onboard the MPO spacecraft, reporting high emission performance (~6 mA, ~0.6 W; Gurnee et al., 2012). The material has extremely low work function (Fig. 1) but is chemically reactive and altered with exposure to gases like water vapor and oxygen, even when cold (e.g., Cronin 1981). In turn, $Y_2O_3$-coated iridium filament which we decided to use is tolerant to the chemically active gases; the material can be applied to many planetary environments.

A $Y_2O_3$-coated iridium disk cathode was applied to the NIM instrument onboard the JUICE spacecraft, also reporting high emission performance (40–400 μA, 1.0–1.3W; Föhn *et al*., 2021), while we used $Y_2O_3$-coated filament cathode. We designed our cathode with higher emission value (20 μA–2 mA; see the following sections). We also note that the filament-type cathode has higher structural robustness against shock/vibration.

## 2.2. Electrical optics designing

We designed a cathode assembly using an iridium filament with an $Y_2O_3$ coating. Fig. 1a shows a cross-sectional view of the surrounding structures. We introduced a beam-focusing Pierce electrode in front of the cathode designed theoretically to obtain the parallel flow of the electron beam (Pierce, 1940). The beam-focusing electrodes have an inclination angle $\theta = 3\pi/8$ rad, known as the Pierce inclination angle. In addition, the suppressor electrode was mounted on the backside of the cathode to compensate for the equipotential line leakage.

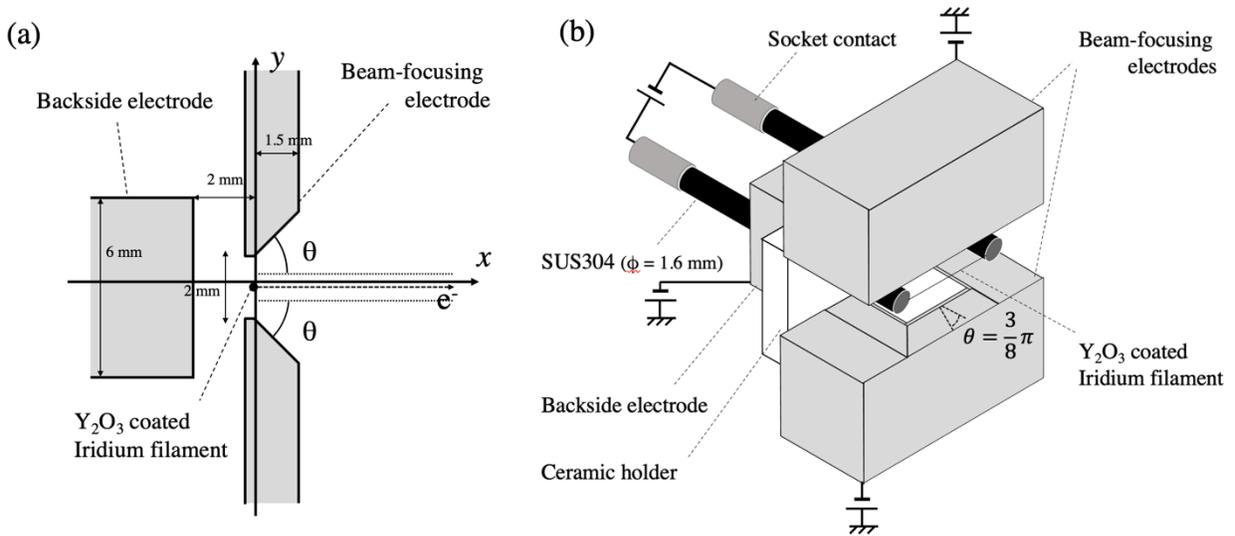

**Fig. 1.** (a) Conceptual two-dimensional diagram of the cathode assembly. (b) Conceptual three-dimensional diagram of the cathode assembly.

Fig. 1b shows a three-dimensional view of the cathode assembly. The ends of the iridium filament were welded to SUS304 pins (1.6 mm diameter), and the length of the filament was 8 mm, including the welding margin. The SUS304 pins constituted the cathode circuit; thus, the ends of the pins were connected to the power supply via socket contacts. SUS304 pins were joined to ceramic by welding to fix the mechanical structure. For the redundancy, two cathode assemblies shown in Fig. 1b were mounted opposite to each other in the instrument (hereafter optics I and II).

Fig. 2a and 2b show numerical simulation results of the instrument for optics I (far side of the exit slit, Fig. 2a) and optics II (near side of the exit slit, Fig. 2b) respectively, with equipotential lines (blue) and beam trajectories (black and red). The cathode assemblies were integrated with electron-beam extraction, electron-trap, and ion-guide electrodes. The electron-beam extraction electrodes were mounted in front of the cathode with a 2-mm pierced hole for the electron-beam paths. The extraction electrodes were connected to a high-voltage supply sufficient to surpass the space-charge limited emission mode, in which the electron current density $J$ [A/m$^2$] was limited by the Child–Langmuir law, as described in equation (2):

$$J = \frac{4\varepsilon_0 V}{9d^2}\sqrt{\frac{2qV}{m_e}} \qquad (2)$$

where $\varepsilon_0$, $V$, $d$, $q$, and $m_e$ denote the vacuum permittivity, voltage difference between the cathode and the extraction electrode, the distance between the cathode and the extraction electrode, elementary charge, and electron mass, respectively. We set the voltage of the extraction electrode to 300 V to surpass $J > 1$ mA/ mm$^2$. The electron capture electrodes were mounted after the electron-beam extraction electrodes, containing holes of the same size as the extraction electrode. The near-side electron capture electrode works as an electron beam lens, and the far-side one works as electron trap that measures the ionizing electron current to calibrate the ionization efficiency. The ion-guide electrodes were mounted to guide and converge the ions toward the ion-beam exit. The voltages of the electron trap and the ion-guide electrodes were set at approximately 70 V to optimize the ionization efficiency, while cross-sections of EI ionization for many gas species were maximal when ~70 eV electrons are bombarded (Engel, 1965).

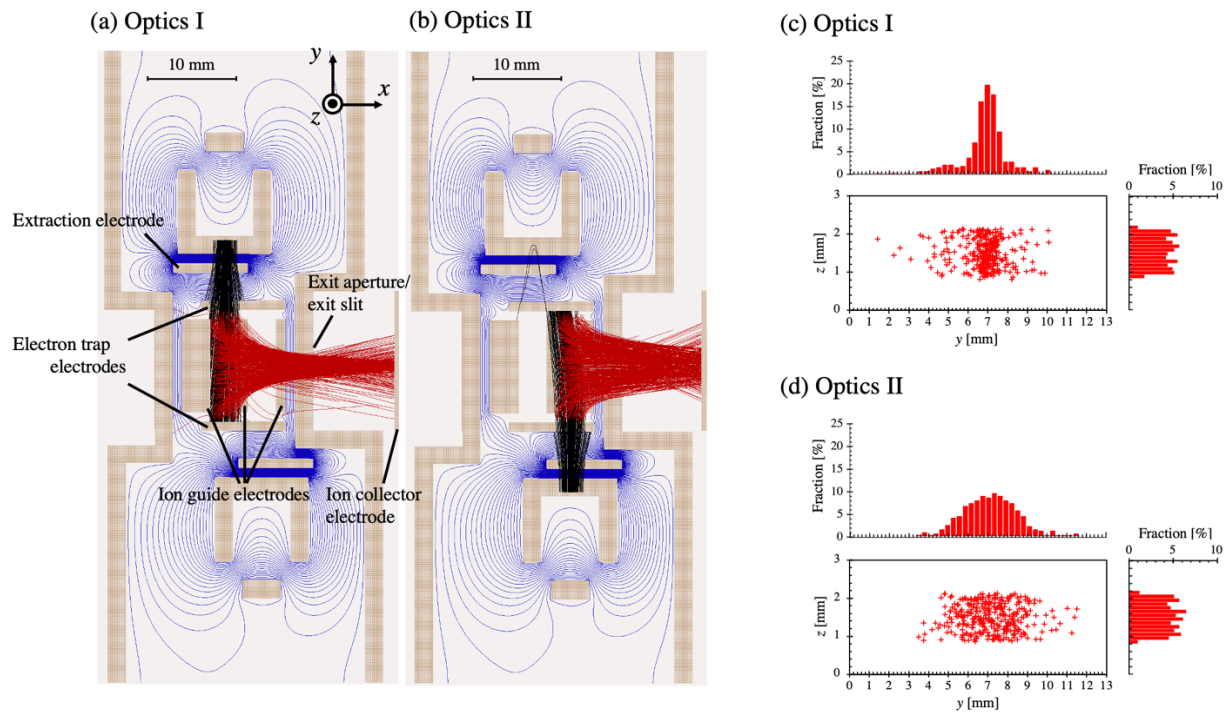

**Fig. 2.** Simulation results of the instrument design. (a) The simulation results with calculated equipotential lines (blue lines), electron trajectories (black lines), and ion trajectories (red lines) for optics I. The name of the electrodes is indicated in the figure. (b) Same results as those of (a) but for optics II. (c) The footprint of ion beams for optics I at the position of ion collector electrode. The distribution of the ion footprints is shown as a histogram for both axes. (d) Same results as those for (c) but for optics II.

In the simulation (Fig. 2), electrons (shown by black lines) started from the surface of the cathodes randomly and ions (shown in red lines) started from the randomized points on the traces of the electron beam, assuming that the electron beam bombardment produced the ions. We established that quasi-parallel flow of the electron beam was achieved and that the ions produced in the ionization area flowed toward the exit slit for both optics I and optics II.

Both Fig. 2c and 2d show the ion footprint distributions for optics I and optics II. We confirmed that ~80% of the ions produced in the ionization region come to the exit aperture (6 mm × 13 mm) and ~40% of those pass through the exit slit (1 mm × 13 mm), resulting in a total ion extraction efficiency $\alpha$ of ~30%. The transmittance for the 1 mm × 13 mm slit was almost the same for both optics I and II, whereas the ion-beam confining effect was more efficient for optics I. Note that the exit slit aperture was determined by the requirements of the reflectron time-of-flight mass spectrometer, which was customized for a particular use.

3. **Results and discussion**

**3.1. Performance test of the ion source**

In this section, we introduce the results of the performance tests in our laboratory. Fig. 3 shows a prototype of the ion source. The size was compact and dimensionally occupied 30 mm × 26 mm × 70 mm, and the optics weighed ~70 g. Electrodes were mounted on a Poly Ether Ether Ketone (PEEK) plate and sandwiched with another PEEK plate.

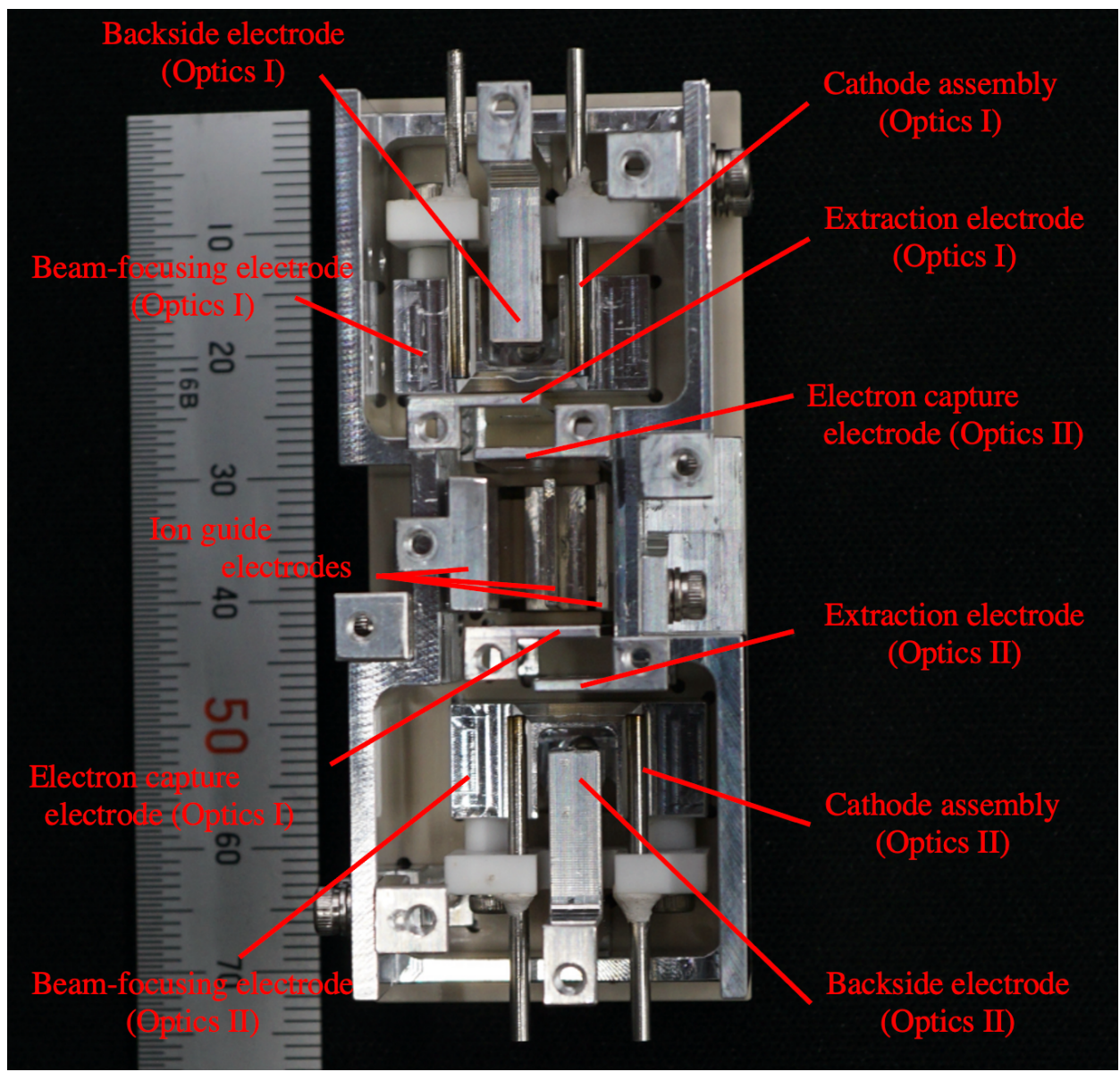

**Fig. 3.** Prototype model used in laboratory experiments. The length unit of the ruler in the picture is a millimeter.

In the following experiment, we used Keithley 6517B pico ammeter/voltage source for current measurement. Using the auto-range option of Keithley 6517B, the measurement errors of the electron/ion currents are <0.1 %. As a filament power supply, we used TAKASAGO Limited

KX-S series DC power supply with 0.01 V CV-resolution. The measurement errors of power consumption are <0.02 W. Additionally, introduction of gas analytes was regulated by CANON ANELVA Corporation ultra-high vacuum variable leak valve, and the analytes gas pressure was monitored by PFEIFFER PBR 260 gauge. Ion current measurements were conducted several minutes after the introduction of analytes gas, confirming the stabilization of environmental pressure level. The measurement errors of the analytes pressure are <5%.

Fig. 4a shows the results of the electron emission experiment. The horizontal axis shows the heating power of the cathode, and the vertical axis shows the amount of electron emission for both optics I and II. The results indicate that extremely little emission occurred below the critical value of the heating power (~2 W). Exceeding the value, the emission level exponentially increased with the growing amount of the heating power (i.e., the temperature of the cathode), presumably obeying equation (1). Above the ~3 W heating condition, the iridium filament will burn out and/or the emission current will be saturated due to the space-charge limitation as described in equation (2). Note that we could not measure the cathode temperature in the experiments because the cathode thickness was below the footprint size of the non-contact thermometer; instead, we verified equation (1) using a calculation method (Fig. S1 in Supplementary Material). We also confirmed that an electron emission exceeding 1 mA could be obtained with our new ion source, which is one order of magnitude larger than conventional space-borne ion sources (Niemann et al., 2002; Balsiger et al., 2007; Mahaffy et al., 2012).

Fig. 4b shows the results of the ionization experiment. In this experiment, the cathode was heated using ~2.8 W to exceed the 1 mA emission level (Fig. 4a). The analyte gas was 99.9% nitrogen, and the background pressure in the vacuum chamber was below $3 \times 10^{-5}$ Pa. The ion current was monitored at the ion collector electrode, mounted after the exit slit. The horizontal axis was the analyte gas pressure introduced to the ion source, and the vertical axis was the amount of ion current ejected from the exit slit. The results showed that the ionization efficiency of the ion source is ~$10^4$ nA/Pa for both optics, which ensured the redundancy of the ion source. We also presented the linearity between ionization efficiency and emission current in the range of 0.1 – 1 mA (Fig. S2 in Supplementary Material), which implies the higher ionization efficiency of our model than that of the conventional models, at least several times.

For EI method, the relationship between the analyte gas pressure $p$ [Pa] and measured ion current $I_i$ [A] can be theoretically described using equation (3):

$$I_i = \alpha \sigma l \frac{p}{k_B T} I_e \qquad (3)$$

where $\sigma$ is the ionization cross-section of the analyte, $l$ is the length of the ionization area, $k_B$ is the Boltzmann constant, $T$ is the temperature of the analyte gas, and $I_e$ is the ionizing electron current. For quantitative analysis, equation (4) can be derived from equation (3):

$$I_i = 3.0 \times 10^{-5} \alpha p \qquad (4)$$

where we assumed $\sigma = 9.5 \times 10^{-20}$ m² for the $N_2$ ionization cross-section by 70 eV electron bombardment (Engel, 1965), $l = 10 \times 10^{-3}$ m, $T = 300$ K as the room temperature, and $I_e = 0.13 \times 10^{-3}$ A referring to the amount of ionizing electron beam measured on the opposite electron-trap electrode. Considering equation (4), we fitted a linear correlation between the analyte pressure and ion current on the experimental data in Fig. 4b, resulting in $\alpha \approx 32\%$. This value of $\alpha$ was quantitatively consistent with the results of numerical calculations (see Section 2.2).

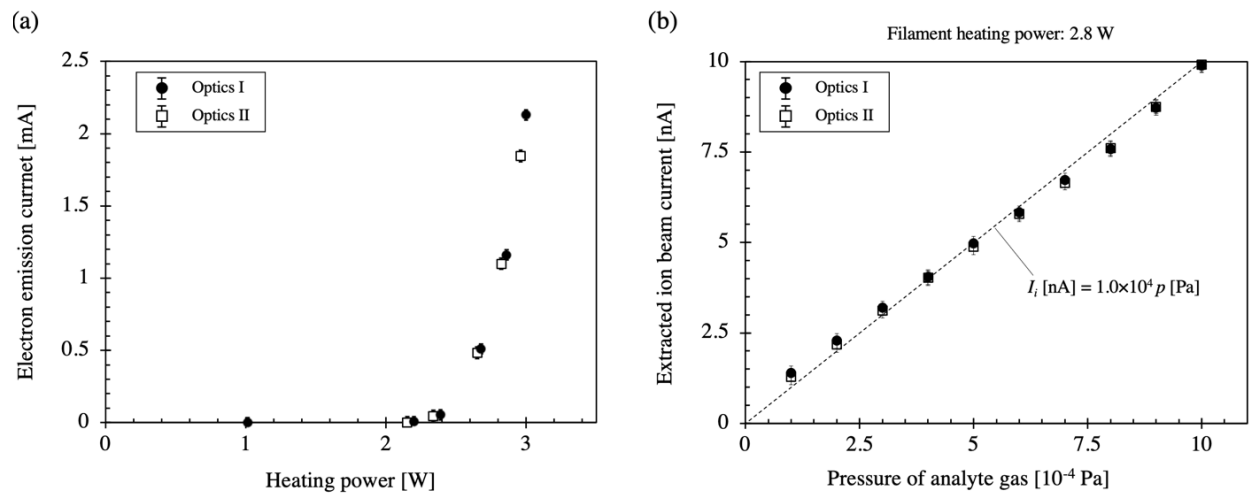

**Fig. 4.** (a) The results of the electron emission experiment. The horizontal axis is the heating power of the cathode, and the vertical axis is the electron emission current of the cathode. The error bars indicate the statistical standard deviation of the electron current measurement. (b) The results of the ion-beam extraction experiment. The horizontal axis is the analytes gas pressure, and the vertical axis is the ion current measured on the ion collector electrode. The error bars indicate the statistical deviation of the ion current measurement. The dashed line shows the best equation (4) fit for the experimental results.

### 3.2. Endurance test of the filament

We performed endurance tests on the filament with repeated operations. In this section, the results of endurance tests are presented.

Fig. 5 shows the block diagram of the test. We used optics II for this experiment. During the test, four physical parameters were monitored: the heating power of the cathode, the ionizing electron current, the ion current, and the analyte gas pressure. The heating power of the cathode was obtained from the value indicated on the power supply. The ionizing electron current was measured at the electron trap electrode opposing optics II. The ion current was monitored at the ion collector electrode after the exit slit. The analyte gas pressure was determined using the ion pressure gauge of the vacuum chamber.

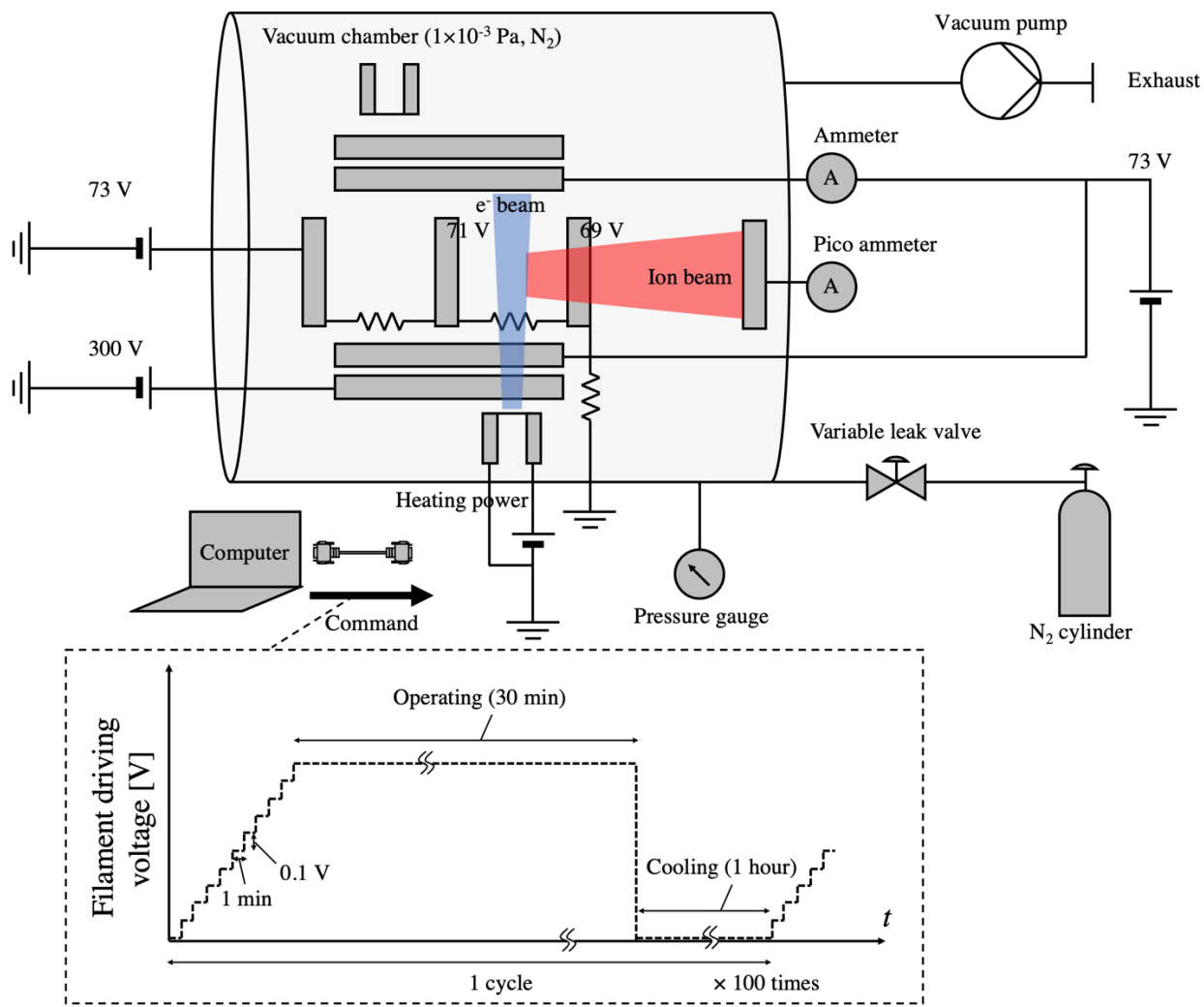

**Fig. 5.** Schematic diagram of the endurance test.

The background pressure in the vacuum chamber was below $3 \times 10^{-5}$ Pa, and the analyte $N_2$ gas pressure in the chamber was maintained at $\sim 1 \times 10^{-3}$ Pa within a 20% fluctuation by trimming a variable leak valve. An external power supply regulated the heating voltage, and the power supply was periodically controlled by commands from a computer. We conducted 100 cycle tests; each cycle was composed of a 10 min heating setup time, 30 min working time, and one hour cooling time. At the beginning of each cycle, the cathode driving voltage was raised to 1.0 $V_{DC}$ with 0.1 V/min increments. Note that the rapid voltage rise to the nominal heating condition was avoided to obviate damaging the cathode. After reaching the nominal heating level (1.0 $V_{DC}$, 2.8 $A_{DC}$), the time responses of the four observables were recorded for 30 min. After the 30 min observation, the driving voltage was switched off.

Fig. 6 compares the time responses of the observables between the first and last cycles. Fig. 6a and 6b show the transition in heating power. The results indicate that the heating power or resistance at the voltage-regulated cathode circuit remained unchanged within 0.1% during 100-times operation. While the resistance of the cathode circuit was affected by the cross-sectional area of the cathode, filament evaporation or the decomposition reacting with environmental gases would appear in the heating power. Thus, the results indicated that the operating temperature and voltage-rise step were adequate for at least 3,000 min or 100 heating cycles.

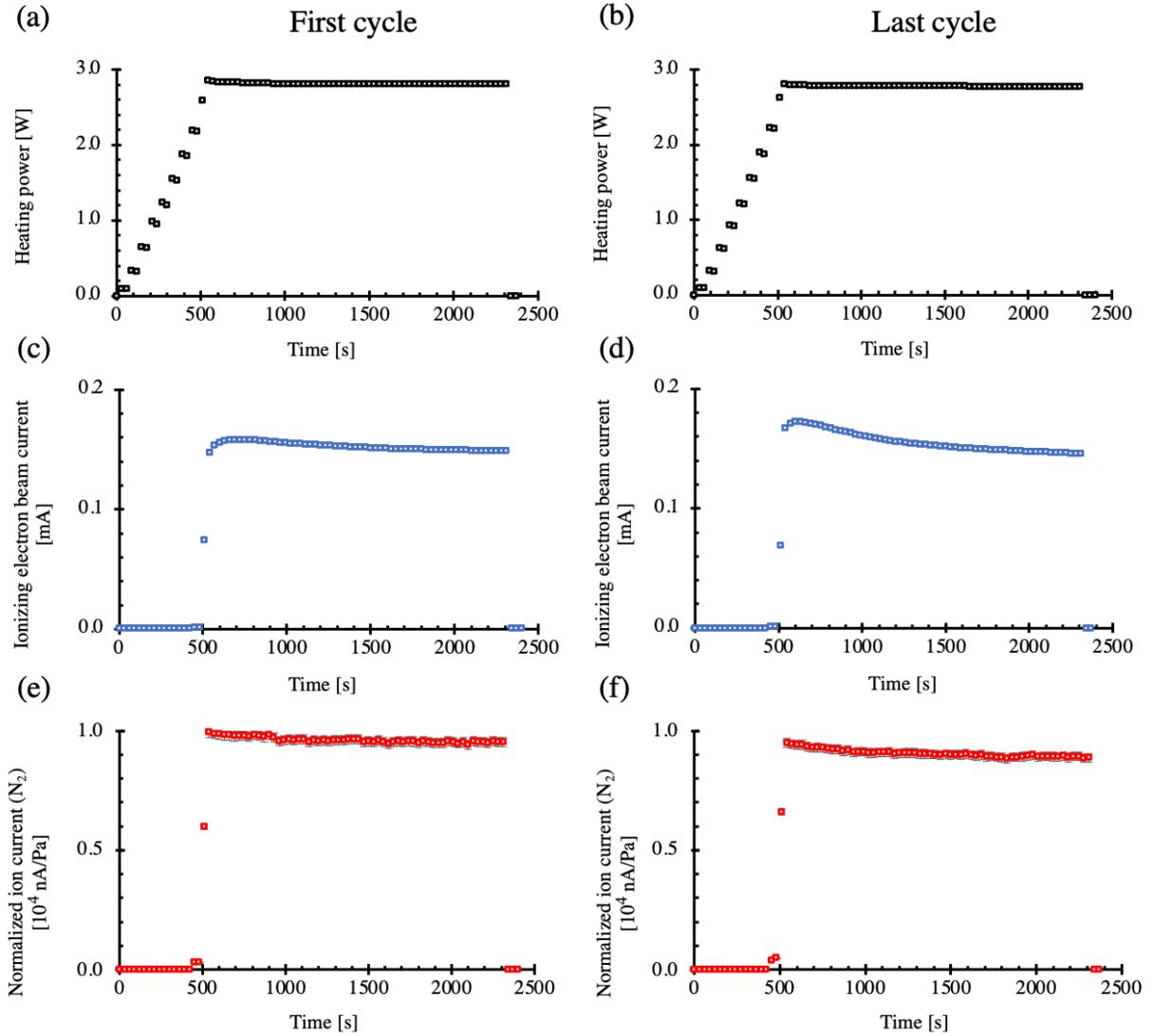

**Fig. 6.** The results of the first cycle and last cycle in the endurance test. The horizontal axis is the time from the start of the cycle. (a) Time series data of the heating power of cathode in the first cycle. (b) Same data as that of (a) for the last cycle. (c) Time series data of the ionizing electron current measured on the ammeter (shown in Fig. 6) connected to the electron-trap electrode. (d) Same data as that of (c) for the last cycle. (e) Time series data of the ion current measured on the pico ammeter (shown in Fig. 6) normalized by analytes pressure. The error bars are derived from the inaccuracy of analytes pressure measurement or minimum unit of the ion gauge. (f) Same data as that of (e) for the last cycle.

Fig. 6c and 6d show the time responses of the ionizing electron current in the first and last cycles, respectively. The results show a transition of the cathode temperature toward a stationary state, indicating that the electron emission was constrained by the temperature regime, according to equation (1). After the heating power reached a nominal value, the ionizing electron current increased until 700 s. This trend can be attributed to the gradual heating of the oxide coating by thermal conduction, which leads to an increase in electron emission from the cathode. Subsequently, the electron emission gradually decreased during the interval 700–2,000 s. This trend can also be reasonably explained because the filament and oxide coating will initially reach the maximum temperature and then gradually cool off due to the heat valance. We realized that this electron-emission overshoot, which may be determined by the $Y_2O_3$ deposition conditions on the iridium filament, changed by 10% after 100-times operation. However, the settling time of the electron emission remained at ~2,000 s, and the settled emission level was consistent within 1%.

Figs. 6e and 6f show the time responses of the ion current normalized by the measured analyte pressure. The purpose of normalization was to cancel the pressure variation effects on the ion current. Each plot in the figures corresponds to the fitting line slope in Fig. 4b. The results showed that a high ionization efficiency (~$10^4$ nA/Pa) was achieved over a period of cycles and even after 100-times operations. We found that the normalized ion current showed a declining trend from 600 s to the end of one cycle, while the ionizing electron current showed an increasing trend ranging from 600–700 s. Considering equation (3), the ion current normalized by the analyte pressure should be proportional to the ionizing electron current; thus, these trends were inconsistent with each other in the range of 600–700 s. This discrepancy may be explained by (1) the initial degassing effect and (2) the temperature profile changes with time at the cathode. For the former explanation, degassing should occur first, enhancing the local pressure around the filament and increasing the ion current. For the latter explanation, the electron beam emitted from the front aspect of the cathode passes through the efficient position for ion extraction, while electrons from other parts of the cathode ionize analytes at inefficient positions. If we assume that the cathode was first heated from the front part, followed by the other inefficient parts, then the electron emission level should change, but the extracted ion current will not. Practically, the ionizing electron current and normalized ion current should be correlated after ~700 s, indicating a gradual heat balance in the cathode circuits.

Fig. 7 shows the normalized ion current over 100 cycles. The horizontal axis is the number of cycles, and the vertical axis is the average value and standard deviation of the normalized ion current in the last minute over 30 mins of operation. We verified the persistency of the normalized ion current, as the value remained constant within 5% throughout the test. We also found that the normalized ion current showed relatively high value in the first five cycles presumably due to the degassing effect around the hot filament.

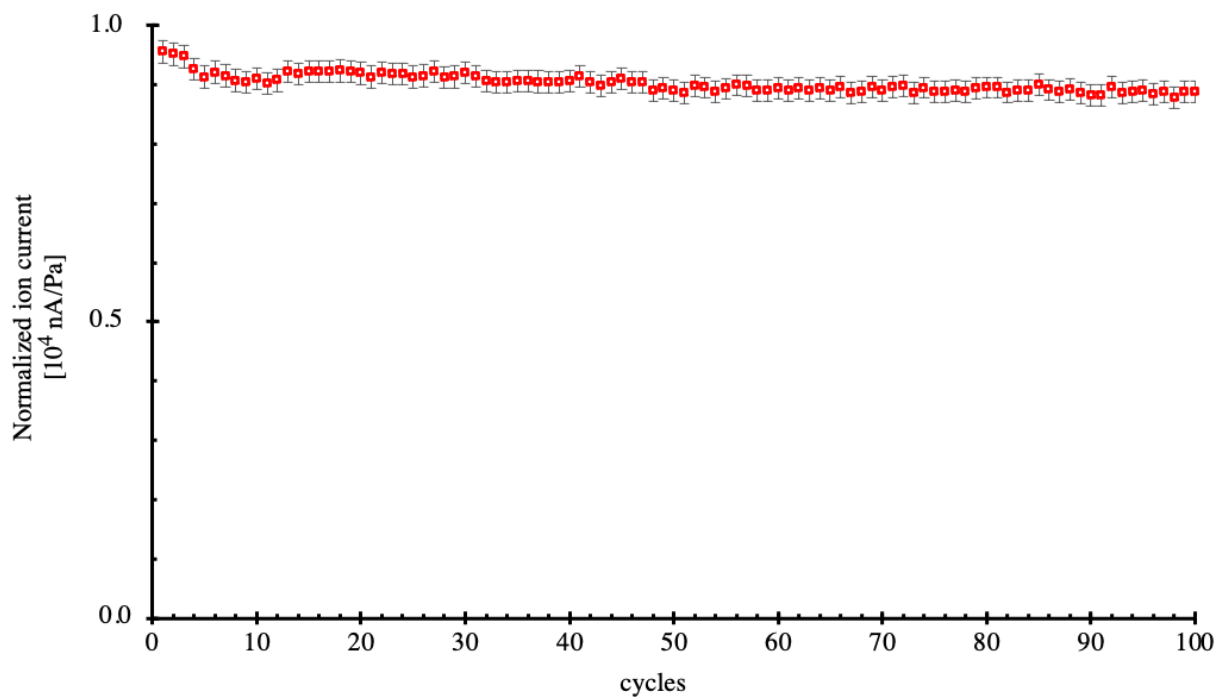

**Fig. 7.** The average value and standard deviation of ion currents in the last 1 min of each cycle. The horizontal axis shows the number of cycles. The vertical axis shows the ion current normalized by analytes pressure. The error bars indicate the statistical deviation of the normalized ion current value during the last 1 min of each cycle.

## 4. Conclusions

We developed a new ion source using the electron impact method to optimize the ionization efficiency, especially for high-sensitivity mass spectrometers required in future planetary missions. In this study, we designed the instrument through simulation and verified its performance using laboratory experiments. The ion source was composed of tandem cathodes using straightened $Y_2O_3$ coated iridium filament, Pierce-type beam focusing electrodes, electron extraction electrodes, and ion-guide electrodes. The main conclusions of this study are summarized as follows:

1. The ion source was developed to have a compact size. The optics dimensionally occupied 30 mm × 26 mm × 70 mm and weighed approximately 70 g. We confirmed that the cathode in the ion source could emit an electron beam ~1 mA with ~2.8 W heating power. As a comparison, the electron gun used as conventional space-borne ion sources were reported to emit 100–200 μA electrons with ~1 W of power. New ion sources can have higher sensitivity by one order of magnitude than the conventional ones.
2. The high ionization efficiency (~$10^4$ nA/Pa) of the ion source was confirmed in the experiment using $N_2$ gas. Additionally, we validated the high ionization performance of both tandem optics, which ensured the redundancy of the ion source.
3. The persistency of the high ionization efficiency throughout 100-cycles of operational tests (for 30 min/1 cycle condition) was observed. We verified the feasibility of a 0.1 V/min increment operation for cathode heating. Additionally, we verified the transient time of the electron emission current (~1,500 s) and adequate starting time for observation after the initial heating operation (~700 s).

**Acknowledgements**

We would like to thank Editage (www.editage.com) for the English language review.

**Funding**

This work was in part supported by JSPS KAKENHI Grant Number 21J10032, 21H04509.

**Supplementary figure**

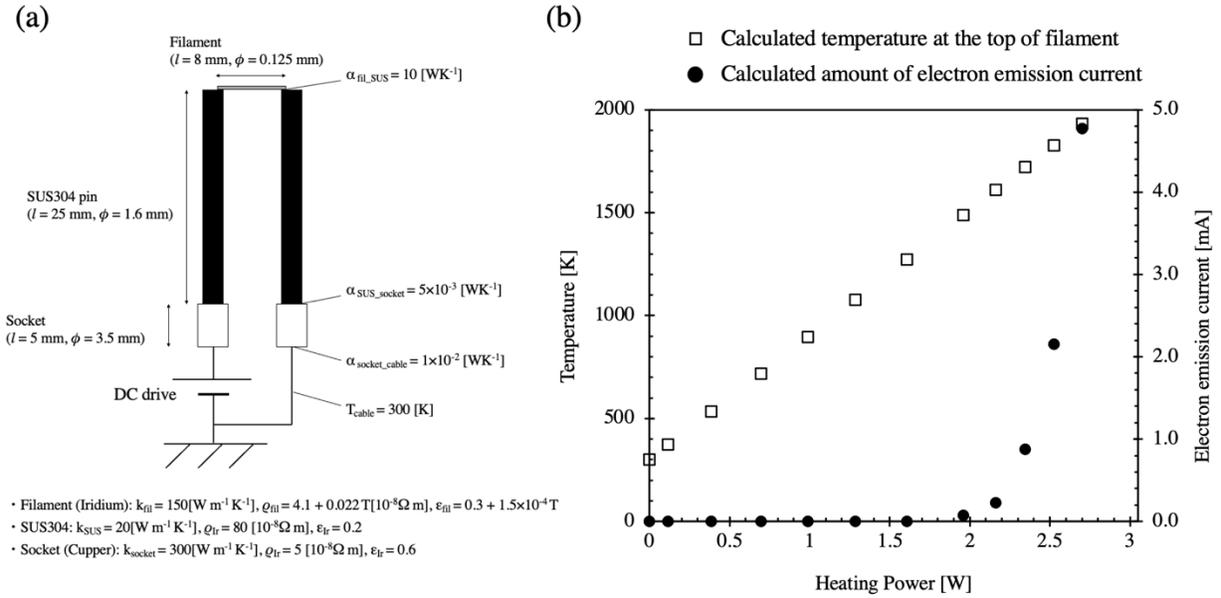

**Fig. S1.** (a) Configuration of the numerical calculation for $Y_2O_3$-coated Iridium cathode with dimensions $\phi=0.125$ mm and $l = 8$ mm. (b) Model dependence of the cathode temperature and emission current as a function of the cathode heating power.

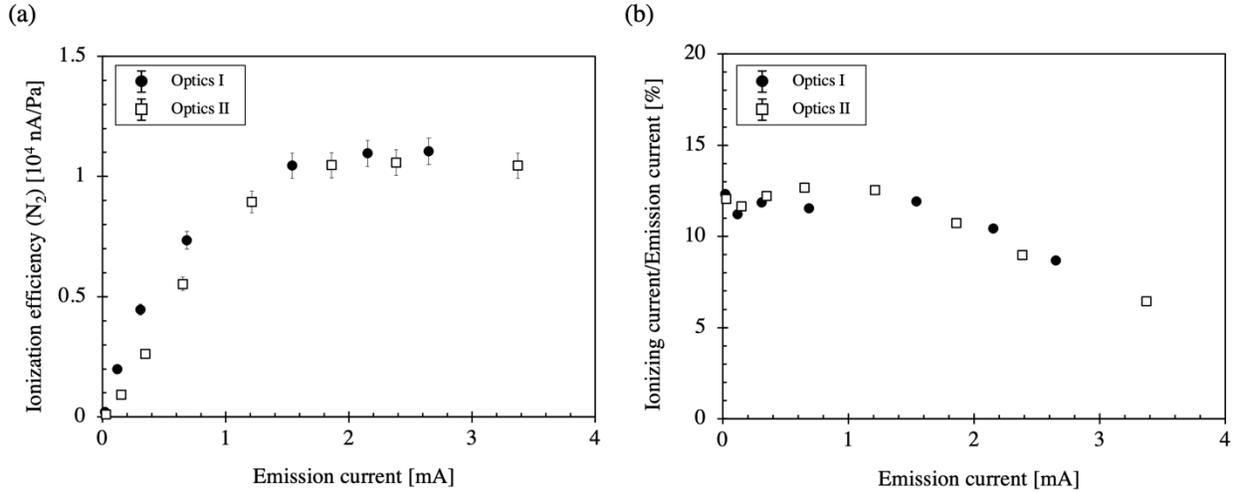

**Fig. S2.** Experimental results showing performances of the ion source as functions of emission. (a) The horizontal axis shows emission current, and the vertical axis shows ionization efficiency. The error bars indicate the errors in measurement. (b) The horizontal axis shows emission current, and the vertical axis shows the fraction of ionizing electron current (measured on the opposite electron-trap electrode) to the total emission current. The error bars indicate the errors in measurement.